\documentclass[twocolumn,usenatbib,useAMS]{mnras}

\usepackage[dvipsnames]{xcolor}

\usepackage[utf8]{inputenc}
\usepackage{geometry,pdfpages}
\usepackage{amsmath}
\usepackage{xfrac}
\usepackage{amsfonts}
\usepackage{amssymb}
\usepackage{natbib}
\usepackage{graphicx}
\usepackage{caption}
\usepackage{subcaption}
\usepackage{framed}
\usepackage{hhline}
\usepackage{float}

\graphicspath{ {./Figures/} }
\newcommand{\Mpc}{{\rm Mpc}}
\newcommand{\kpc}{{\rm kpc}}

\newcommand{\Msolar}{{M_{\odot}}}
\newcommand{\AEST}{AeST}

\newcommand{\grad}{\ensuremath{\vec{\nabla}}}

\newcommand{\ah}{a_h}

\newcommand{\chib}{{\bar{\chi}}}

\newcommand{\lamdimless}{\hat{\chi}_{\rm out} }
\newcommand{\lamdimlessmax}{\hat{\chi}_{\rm out}^{(max)}  }

\newcommand{\chiin}{\chi_{\rm in}}
\newcommand{\chiout}{\chi_{\rm out}}

\newcommand{\lambdas}{\lambda_s}

\newcommand{\Phih}{\tilde{\Phi}}

\newcommand{\Phit}{\tilde{\Phi}}

\newcommand{\Ycal}{{\cal Y}}

\newcommand{\Jcal}{{\cal J}}

\newcommand{\GN}{\ensuremath{G_{{\rm N}}}}

\newcommand{\rchi}{\ensuremath{r_{\chi}}}
\newcommand{\rint}{\ensuremath{r_{I}}}
\newcommand{\rM}{\ensuremath{r_{{\rm M}}}}
\newcommand{\rMh}{\ensuremath{\hat{r}_{{\rm M}}}}

\newcommand{\rC}{\ensuremath{r_{{\rm C}}}}

\usepackage{wrapfig}
\usepackage{hyperref}

\usepackage{cuted}  

\usepackage{ulem}  
\def\bi{\begin{itemize}}
\def\ei{\end{itemize}}

\title[{\AEST}: Quasistatic spherical solutions]{Aether Scalar Tensor ({\AEST}) theory: Quasistatic spherical solutions and their phenomenology.}
\author[P. Verwayen et al.]{
Peter Verwayen$^{1}$ \thanks{E-mail: peter.verwayen@sydney.edu.au} 
Constantinos Skordis$^{2}$
and C\'eline B\oe hm$^{1}$
\\
$^{1}$School of Physics, Faculty of Science, University of Sydney, NSW 2006, Australia \\
$^{2}$CEICO - FZU, Institute of Physics of the Czech Academy of Sciences, Na Slovance 1999/2, Prague 182 000, Czechia
}

\date{Accepted XXX. Received YYY; in original form ZZZ}

\pubyear{2023}

\hypersetup{draft}

\begin{document}
\label{firstpage}
\pagerange{\pageref{firstpage}--\pageref{lastpage}}
\maketitle

\begin{abstract}
There have been many efforts in the last three decades to embed the empirical MOND program into a robust theoretical framework. While many such theories can explain the profile of galactic rotation curves,  they usually cannot  explain the evolution of the primordial fluctuations and the formation of 
large-scale-structures in the Universe. The Aether Scalar Tensor (AeST) theory seems to have overcome this difficulty, thereby  providing the first compelling example of an  extension of general relativity able to successfully challenge the particle dark matter hypothesis. 
Here we study the phenomenology of this theory in the quasistatic weak-field regime and specifically for the idealised case of spherical isolated sources. 

We find the existence of three distinct gravitational regimes, that is, Newtonian, MOND and a third regime characterised by the presence of oscillations in the gravitational potential which do not exist in the traditional MOND paradigm. 
We identify the transition scales between these three regimes and discuss their dependence on the boundary conditions and other parameters in the theory.
Aided by analytical and numerical solutions, we explore the dependence of these solutions on the theory parameters.
Our results could help in searching for interesting observable phenomena at low redshift pertaining to galaxy dynamics as well as lensing observations,
however, this may warrant proper N-body simulations that go beyond the idealised case of spherical isolated sources.
\end{abstract}



\section{Introduction}

The presence of an invisible  matter -- dark matter -- permeating the Universe throughout its cosmic evolution is the most popular explanation 
to the origin and number of large-scale-structures in the Universe as well as their internal dynamics. 
Yet dark matter particles remain elusive despite the vast array of experimental programmes searching for such particles with great sensitivity, see e.g.~\cite{Agrawal:2021dbo,LZ:2022ufs,XENON:2023sxq}.  

While the particle dark matter hypothesis is still far from being excluded, the task of coming up with alternative theories which fit galactic and cosmological observations 
remains important and becomes increasingly so, especially in the absence of direct evidence for particle dark matter.
One of such alternatives, namely Modified Newtonian Dynamics (MOND), was formulated in~\cite{Milgrom83} and~\cite{1984ApJ...286....7B} to explain
 the anomalous rotation curves of galaxies at large radii. MOND postulates that Newton’s second law must be modified for accelerations $\vec{a}$ with magnitude
smaller than $a_0  \sim 1.2 \times 10^{-10}$m$/$s$^2$ to explain the rotation curves of galaxies \citep{Begeman_Broeils_Sanders_1991, 1988ApJ...333..689M}. This is neatly captured by changing  
Newton’s second law to $\mu(y) \vec{a} = -\nabla \Phi$,  with $y\equiv|\vec{a}|/a_0$ and $\Phi$ being the gravitational potential determined from the standard Poisson equation sourced by baryons only.
This works if $\mu(y) \simeq y$ for $y<1$ (this is the so-called MOND regime) 
and $\mu(y) = 1$ for $y>1$ (this is the Newtonian regime).

The MOND proposal has had several successes such as \citep{Milgrom1988,Kent1988,Begeman_Broeils_Sanders_1991,Sanders:1996ua,Sanders:2018mnk,McGaugh2020,McGaugh:2000sr,McGaugh:2005qe,LelliEtAl2019,McGaughEtAl2016} 
 which have kept it alive~\citep{2020pama.book.....M}. 
However, challenges remain, as it appears to be in conflict with observations of galaxy clusters~\citep{Sanders1999,Sanders:2002ue,Pointecouteau:2005mr,FamaeyMcGaugh2011,Ettori:2018tus,Tian:2020qjd}, 
the bullet cluster \citep{2006ApJ...652..937B, 2006ApJ...648L.109C}, and dwarf spheroidal galaxies \citep{GerhardSpergel1992,Serra:2009tj,AlexanderEtal2017,Read:2018fxs}.

MOND faces a fundamental theoretical problem: it is not a relativistic theory but rather an empirical model.
As such, MOND cannot be used to compute the formation and distribution of large-scale-structures in the Universe. 
A number of relativistic theories, typically extensions of General Relativity (GR), have been proposed to palliate this issue,
 and all have been constructed to lead to MOND behaviour in the non-relativistic limit (see \cite{famaey_modified_2012} for a review). 

One of these theories, the Tensor-Vector-Scalar (TeVeS) theory~\citep{Sanders_1997-ApJ480,Bekenstein_2004-PhRvD_70h3509B},  introduces a unit-timelike vector and a scalar field in addition to the usual gravitational metric tensor. These fields are combined together to define a second metric tensor which is used
to determine the geodesics of ordinary standard model matter. With this bi-metric disformal (owing to the vector field) structure, TeVeS leads to the equality of dynamical and lensing mass and thus
can generate the right amount of gravitational lensing produced by baryon-only galaxies as if dark matter was present. Perturbations of the vector field also play a major role cosmologically
as their time-evolution can lead to matter power spectra in line with observations~\citep{Skordis:2005xk,Dodelson:2006zt}. However, despite these important improvements  over other theories,
TeVeS fails to fit the angular distribution of the CMB~\citep{Skordis:2005xk}. Moreover, TeVeS leads to tensor mode gravitational wave speed $c_{GW}$ which is different than the speed of light
$c_{EM}$ and thus has been ruled out by the simultaneous observation of the GW170817 and GRB170817A events\citep{LIGOScientific:2017zic}.

The recent Aether Scalar Tensor (\AEST) proposal of \cite{Skordis_Zlosnik_2021} retains the unit-timelike vector field and the scalar from TeVeS,
but only has one metric tensor and thus, no bi-metric structure. Unit-timelike vector fields have been dubbed ``aether fields'' in other instances, see~\cite{Jacobson:2000xp},
hence, the naming of this theory\footnote{The name {\AEST} does not appear in the \cite{Skordis_Zlosnik_2021} but only in the followup article \cite{Skordis:2021mry}.}.  
The success of {\AEST} theory rests on maintaining $c_{GW} = c_{EM}$ in all situations and fitting the CMB and matter power spectrum data quite convincingly, 
 while retaining a MOND limit in galaxies and the correct gravitational lensing.

Cosmologically, the scalar in {\AEST} theory evolves as in shift-symmetric k-essense~\citep{Scherrer:2004au}  which results in its cosmological energy density being similar to dust, i.e. $\propto (1+z)^{3}$ plus small decaying corrections. This k-essence-like behaviour of {\AEST} theory leads to spontaneous breaking of time diffeomorphisms as in the Ghost condensate (GC) theory~\citep{Arkani-Hamed:2003pdi}
which results in the metric potential $\Psi$ (see \cite{Skordis_Zlosnik_2021,Skordis:2021mry} for details) acquiring a mass term $\mu$. The quasi-static long-distance behaviour of {\AEST} theory 
thus departs from MOND, but is still different than the GC due to the presence of the non-canonical MOND term and the vector field. Our aim in this paper is to precisely understand 
the particular behaviour of {\AEST} theory in spherically symmetric static situations, due to the new features introduced by $\mu$.

We present the {\AEST} theory in Section~\ref{Theory}, and derive the gravitational equations for quasistatic distributions of matter. 
In Section~\ref{Spher_sym}, we study the properties of density profiles assuming spherical symmetry. This work is critical in view of
 creating N-body simulations of large-scale-structure formation in the late Universe to compare \AEST~theory with non-linear observables, such as, those related to galaxy clusters (including the 
bullet cluster). It can also provide a guide on how extensions of GR at both early and late times may be distinguied from models based on the dark matter hypothesis.

\section{Theoretical Model}
\label{Theory}

We now summarise the main features of {\AEST}~theory as proposed in \cite{Skordis_Zlosnik_2021}, the theory parameters, and
the transition scales between the three regimes (Newton, MOND and ``$\mu$-domination'') which feature in this theory. 
More details regarding the underlying setup and derivation of the weak-field static equations may be found in \cite{Skordis_Zlosnik_2021,Skordis:2021mry}.  

\subsection{Static weak-field equations}

The fields of {\AEST}~theory are the metric $g_{\mu \nu}$, a unit-timelike vector field $A_\mu$, such that $A^\mu A_\mu = -1$, and a scalar field $\phi$.
Perturbing the field equations around Minkowski spacetime and ignoring time derivatives reduces the field dependence to just two potentials: $\Phi$ and $\chi$, see \cite{Skordis_Zlosnik_2021,Skordis:2021mry}.
The gravitational potential $\Phi$ results from the metric perturbation. Matter fields couple minimally to this potential and follow its geodesics such that the Einstein equivalence principle holds.
Specifically, gravitational accelerations are determined from $\grad\Phi$. 
The potential $\chi$ is a gauge-invariant combination of a scalar mode contained~\footnote{The $0$-th 
component of the vector field is fixed due to the unit-timelike constraint while its spatial part is given in general as a gradient of a scalar mode 
and the curl of a spatial vector.  In this study, we ignore the curl mode since it vanishes in spherical symmetry.} 
in the spatial part of the vector field $A_\mu$, and the perturbation of the scalar field $\phi$.
In addition to $\Phi$ and $\chi$, we find it useful to define the potential $\Phih$ through
\begin{align}
    \Phi = \Phih + \chi,
    \label{Phih_Phi_chi}
\end{align}
treating $\Phih$ and $\chi$ as the fundamental fields and $\Phi$ as a derived variable.
With these variables, the weak-field equations of {\AEST}~thaory which extend the standard Poisson equation for $\Phi$ and determine spatial dependence of the gravitational potentials of the theory  $\Phih$, $\chi$ and $\Phi$,
can be written in the form
\begin{align}
    \grad^2\Phih +   \mu^2  \Phi =& \frac{4\pi \GN}{1+\beta_0} \rho_b,
    \label{Phi-Orig}
    \\
    \grad^2  \Phih   =& \grad \cdot \left[\frac{d\Jcal}{d\Ycal} \grad \chi\right],
    \label{Chi-Orig}
\end{align}
while a third, non-independent equation, is
\begin{align}
    \grad \cdot \left[\frac{d\Jcal}{d\Ycal} \grad \chi\right] +  \mu^2  \Phi =& \frac{4\pi \GN}{1+\beta_0} \rho_b.  
    \label{Chi-AQUAL}
\end{align} 
In the equations above, $\GN$ is Newton's constant, $\rho_b$ is the baryonic density, $\Jcal = \Jcal(\Ycal)$ is a function of the variable $\Ycal= (g^{\mu\nu} + A^{\mu} A^{\nu})\nabla_\mu \phi \nabla_\nu \phi$,
$\mu$ is the mass parameter and $\beta_0$ is the inverse screening parameter, the screening parameter being $\lambdas \equiv 1/\beta_0$. The latter determines how much $\chi$ contributes to $\Phi$ through \eqref{Phih_Phi_chi} in the large gradient limit of \eqref{Phi-Orig},
\eqref{Chi-Orig} and \eqref{Chi-AQUAL}; see below. 
The mass parameter $\mu$ is determined through a specific combination of parameters which enter the action of the theory, see \cite{Skordis_Zlosnik_2021,Skordis:2021mry} for details,
however, for our purposes we treat it as a free parameter in this work.  It can be shown that linear stability of the theory on Minkowski space imposes $\lambdas>0$ and $\mu^2>0$, however, 
in practice  $\mu^{-1} \gtrsim 1 \ \Mpc$ to ensure that MOND solutions are attainable at galactic scales. 
Indeed, precise observations of the extent of flat galactic rotation curves can be used to put strong constraints on $\mu^{-1}$ \citep{Mistele:2023paq}.

The variable $\Ycal$ denotes the ``spatial gradient'' combination $\Ycal= (g^{\mu\nu} + A^{\mu} A^{\nu})\nabla_\mu \phi \nabla_\nu \phi$ and in the static weak-field case one finds that
\begin{align}
\Ycal  \rightarrow  |\grad \chi|^2
\end{align}
The function $\Jcal$ then controls the behaviour of  static weak field configurations through its dependence on $\Ycal$, or more specifically on $\sqrt{\Ycal} =  |\grad \chi|$. 
When  $\sqrt{\Ycal} \gg a_0$ the function  $\Jcal$ is constructed so as to recover GR as close as possible. This can happen if 
\begin{align}
\Jcal \rightarrow \lambdas \Ycal \qquad  \text{when} \quad \sqrt{\Ycal} \gg a_0, 
\label{Jcal_GR}
\end{align}
which also serves as the definition of the screening parameter $\lambdas$, a parameter which must be part of the function $\Jcal$. As $\lambdas\rightarrow \infty$ (equivalently $\beta_0\rightarrow 0$), then 
$\chi \rightarrow 0$ so that the contribution of $\chi$ to $\Phi$ through \eqref{Phih_Phi_chi} is totally screened\footnote{Note that this applies only in the $\sqrt{\Ycal} \gg a_0$ limit, that is,
the weak gradient limit is unaffected and $\chi$ still plays a role there.}.

In the opposite limit, $\sqrt{\Ycal} \ll a_0$, the function $\Jcal$ is constructed to lead to MOND behaviour, which is possible if
\begin{align}
\Jcal \rightarrow   \frac{2\lambdas}{3(1+\lambdas) a_0}  \Ycal^{3/2} \quad  \text{when} \quad \sqrt{\Ycal} \ll a_0. 
\label{Jcal_MOND}
\end{align}

Three comments are in order: 
\begin{itemize}
\item In deriving \eqref{Phi-Orig}-\eqref{Chi-AQUAL} one also finds that the two standard  weak-field metric potentials $\Phi$ and $\Psi$ are equal. This ensures that the lensing mass is equal to the dynamical mass. This means that  for any source configuration where the MOND limit of {\AEST}~theory successfully replaces dark matter, the gravitational lensing signal for the same source will also be as if dark matter is present.
\item The case $\chi=0$ corresponds to having $\Phih = \Phi$ and the above system of equations reduces to a single equation for $\Phi$ which is identical to the non-relativistic weak-field limit of the Ghost condensate model \citep{Arkani-Hamed_2004}.
\item Because of the term $\mu^2 \Phi$, the system of equations \eqref{Phi-Orig}-\eqref{Chi-AQUAL} is not precisely that of TeVeS theory but tends to it as $\mu\rightarrow 0$. 
However, dust-like solutions in cosmology require $\mu$ to be non-zero, with $\mu\rightarrow \infty$ corresponding to the pure Higgs phase~\citep{Arkani-Hamed_2004} where the solutions are exactly dust-like throughout the 
entire history of the Universe.  Thus {\AEST}~theory will have solutions which depart from the MOND paradigm at a large enough distance away from the matter source.  This results in a third regime, different from both GR and MOND as we study thoroughly below.
\end{itemize}

\subsection{Newtonian, MOND and \texorpdfstring{$\mu$}{\emph{u}}-domination limits and corresponding transition scales} 
\label{section:Limits_transition_scales}

We define the dimensionless variable
\begin{align}
x  \equiv\frac{|\grad\chi|}{a_0}
\end{align}
and the new function 
\begin{align}
f(x)  \equiv \frac{d\Jcal}{d\Ycal}
\end{align}
which corresponds to the usual MOND interpolation function and is designed to provide a continuous transition between the Newtonian and MOND regime as follows:  
\begin{equation}
\label{def_f_general}
f \rightarrow 
\begin{cases}
    \lambda_s ~~~&  \text{: large gradient limit, } x\gg 1 
\\
    \frac{x}{1+\beta_0} ~~~& \text{: small gradient limit, } x \ll 1 
\end{cases}.
\end{equation}
with $\beta_0 \equiv 1/\lambdas$ (see also the equivalent limits in terms of $\Jcal$ in equations \eqref{Jcal_GR} and  \eqref{Jcal_MOND}).

We note that the large gradient limit corresponds to Newtonian behaviour to lowest order in the potentials and GR-like behaviour in the full relativistic strong-field case (up to parametrically small post-Newtonian corrections).

\subsubsection{Large gradient limit:  strong-field/Newtonian regime}

Consider first the large gradient limit which leads to the strong-field (and Newtonian) regime. Setting $f = \lambdas$ in equation \eqref{Chi-Orig} and integrating twice, results in $\chi  = \beta_0 \Phih + \chi_0$, 
where we have neglected a possible curl which is not relevant in spherically symmetric situations (see also the discussion in \ref{section:analytic_equations}). 
Substituting this relation in \eqref{Phih_Phi_chi} gives $\Phih  = (\Phi-\chi_0)/(1+\beta_0)$
and with these conditions \eqref{Phi-Orig} takes the form of the inhomogeneous Helmholtz equation
\begin{align}
  \grad^2\Phi +  (1+\beta_0)  \mu^2  \Phi =& 4\pi \GN  \rho_b.
  \label{eq:helmholtz}
\end{align}
Equation \eqref{eq:helmholtz} also results from taking the non-relativistic weak-field limit of the Ghost condensate model \citep{Arkani-Hamed_2004}.
For scales smaller than  $\sim\mu^{-1}/\sqrt{1+\beta_0}$, \eqref{eq:helmholtz} reduces to the standard Poisson equation describing Newtonian gravity.
The solution to \eqref{eq:helmholtz} for scales larger than $\sim\mu^{-1}/\sqrt{1+\beta_0}$ is not physically relevant to the models studied in this article, 
since the systems we study first transition to the small gradient regime before the scale $\sim\mu^{-1}/\sqrt{1+\beta_0}$ is reached. 

Note that the integration constant $\chi_0$ does not affect \eqref{eq:helmholtz}, and so, does not play a role in the large gradient regime.
However, it's value will survive until the small gradient regime, and as we discuss below, it will lead to observable effects at very large distances from the source.

\subsubsection{Small gradient limit: MOND and \texorpdfstring{$\mu$}{\emph{u}} regimes}
\label{small_gradient_limit}

Consider now the small gradient limit where $f = x / (1+\beta_0)$; see \eqref{def_f_general}. Then \eqref{Chi-AQUAL}  becomes
\begin{align}
    \grad \cdot \left[ (1+\beta_0) \  f(x) \ \grad \chi\right] +  (1+\beta_0) \mu^2  \Phi =& 4\pi \GN \rho_b,
    \label{eq:small_gradient_fx}
\end{align}
which reduces to 
\begin{align}
    \grad \cdot \left[ \frac{|\grad\chi|}{a_0} \grad \chi\right] +  (1+\beta_0) \mu^2  \Phi =& 4\pi \GN \rho_b,
    \label{eq:small_gradient}
\end{align}
leading to the MOND equation for $\chi$ in the limit where $\mu\rightarrow 0$  and Eq.~\eqref{Phi-Orig} turns into 
\begin{equation}
    \grad^2 \Phih = \frac{4\pi \GN}{1+\beta_0} \rho_b.  
\end{equation}
The above equation has an exterior solution scaling as $|\grad \Phih| \sim 1/r^2$ while the deep MOND solution to \eqref{eq:small_gradient} (for $\mu=0$) is $|\grad \chi| \sim 1/r$.
Hence, for large enough $r$ (but not too large so that we keep within the regime where $\mu$ has no influence and can be set to zero)
we have $|\grad \Phih| \ll |\grad \chi|$, that is $|\grad \Phi| \sim |\grad \chi| \sim 1/r$ implying that the solution for $\Phi$ is also MOND-like.

When $r$ becomes even larger, the $ (1+\beta_0) \mu^2  \Phi$ term in the above equation becomes important, leading to a new regime specific to {\AEST}~theory which departs from MOND. 
This happens because since $|\grad\chi|\sim 1/r$, the 1st term is expected to scale as $1/r^3$, while the 2nd term grows as $\ln r$ and is always bound to become important at some scale $\rC$.
We call this the $\mu$-dominated regime. 

We now derive the transition scales from Newton to MOND and from MOND to  $\mu$-domination in the case of a point source.

\subsubsection{Transition from Newtonian to MOND regime}

The transition from Newtonian to MOND behaviour in the “classical” MOND paradigm occurs when the Newtonian force is equal to the MOND force, and is given by the familiar MOND radius equation
\begin{align}
    \rM  \sim \sqrt{\frac{\GN M}{a_0}},
    \label{eq:r_m}
\end{align}
However, in \AEST~theory, the MOND force is coupled only to the $\chi$ component of the field. Therefore, determining the radius at which the total force deviates from the Newtonian force is determined by the $\chi$ component only.  In the Newtonian limit of \AEST~theory, $\chi = \beta_0 \Phi /(1+\beta_0)$ 
 with $\grad \Phi$ tracking the Newtonian force (i.e.  $|\grad \Phi| = G_N M / r^2$). Hence $\grad \chi$ departs from the Newtonian limit and enters the MOND regime 
when $\grad \chi$ is no longer is proportional to $|\grad \Phi| = G_N M / r^2$. This happens when the two limiting situations in \eqref{def_f_general} become comparable, 
that is, when $\lambdas \sim x / (1+\beta_0)$. Hence, different transition scales may occur depending on whether we consider interior or exterior solutions. 
In the exterior case, we call this transition $r_{\chi}$ where
\begin{equation}
    \rchi = \frac{\rM}{1+ \lambdas},
    \label{eq:r_chi}
\end{equation}
such that $\rchi \le \rM$ always.  Given Eq.\eqref{eq:r_chi} and the fact that $\beta_0 \ll 1$, we expect $r_{\chi} \ll r_M$.

To determine when the total {\AEST}~force enters the MOND regime, we also need to require that $\chi$ is the dominant component of $\Phi$, that is $\grad \Phi \simeq \grad \chi$. 
This domination occurs at a distance $\hat{r}_M$ which is defined as the point when $|\grad\chi| = |\grad\tilde\Phi|$, leading to
\begin{align}
    \rMh  \sim \frac{\rM}{1+\beta_0},
    \label{eq:r_m_h}
\end{align}
such that $\rMh \le \rM$ always. 
In practice, $\beta_0$ is expected to be small so that the two scales $\rM$ and $\rMh$ are approximately equal.

The different transition scales above are depicted in Fig.\ref{fig:NS_l-m1_TH_Hyp} where we discuss how the solutions change when we vary the parameter $\lambdas$.

\subsubsection{Transition from MOND to \texorpdfstring{$\mu$}{\emph{u}}-domination}
\label{Mond_to_mu}

To estimate the transition between MOND and $\mu$-domination we consider \eqref{eq:small_gradient} for a point source of mass $M$ which we define to occur at the scale $r_C$.
Estimating derivatives as $d/dr \sim 1/r$ and considering the Log terms appearing in the MOND solution $\Phi$, i.e. $\ln(r/\rMh) $, as being $O(1)$\footnote{We want to have a MOND regime extending for a few
radii larger than $\rMh$ before transitioning to a $\mu$ regime. Due to the Log, even a few orders of magnitude difference will still give an $O(1)$ number in the grand scale of things.},
the terms on the LHS in \eqref{eq:small_gradient} become comparable at the scale $\rC$ when
\begin{align}
    \frac{1}{r_C^3 a_0} \chi^2 \sim   (1+\beta_0)  \mu^2 \Phi
\label{simple_r_C_estimation}
\end{align}
where $\chi\sim\Phi \sim \sqrt{\GN M a_0} \times O(1)$. This gives a simple estimate  of $\rC  \sim  \left( \rMh /\mu^2 \right)^{1/3}$.
    
The above estimate, however,  ignores possible effects coming from the boundary condition and these can be important.  
Let us set the boundary condition for $\chi$ at a radius $r_0$, that is, $\chi(r_0)$. For later convenience, we  normalize it with respect to $\sqrt{\GN M a_0}$ 
and define $\lamdimless \equiv   \chi(r_0) / \sqrt{\GN M a_0} $  as the constant  free parameter setting the boundary condition.
Consider now the extreme case where $\lamdimless$  is so large that it dominates the solution, while $\grad\chi$ still retains a (subdominant) $1/r$ MOND component. 
Then, $\Phi \sim \chi \sim \sqrt{\GN M a_0} \lamdimless$ and \eqref{simple_r_C_estimation} leads to $\rC \sim \left( \rMh /   \mu^2   \lamdimless   \right)^{1/3} $. 

The last relation above would appear to diverge when  $\lamdimless\rightarrow0$, but nothing obviously wrong should happen when setting the boundary condition for $\chi$ to 
zero (and this is verified numerically). Moreover, the point $r_0$ where the boundary condition is set is arbitrary, thus the ``zero point'' 
where $ \chi(r_0)$ (or equivalently  $\lamdimless$  ) is zero is also arbitrary.

Nevertheless, defining $\lamdimless$ can be done in an unambiguous way, so that a better estimate for $\rC$ which interpolates between the two extreme cases above can be derived.
We summarise this here and leave the details for appendix-\ref{Appendix_rC}.
In the small gradient regime, see \ref{small_gradient_limit}, the details of the interpolation function are unimportant and we may set $f\rightarrow x/ (1+\beta_0)$ from \eqref{def_f_general}
resulting in \eqref{eq:small_gradient}.
First, assuming that $\mu=0$ the solution is  $\chi =  \sqrt{\GN M a_0} \left( \lamdimless + \ln \frac{r}{\rMh}\right)$, which serves to define $\lamdimless$
as $\lamdimless \equiv \chi(\rMh) / \sqrt{\GN M a_0}$.
To determine when the full $\mu\ne0$ solution deviates from the pure MOND solution, we expand the former in terms of the MOND solution plus a small perturbation.
We find that when $\lamdimless$ takes a specific value that we denote as $\lamdimless^{(max)}$, the transition scale $\rC$ reaches a maximum. In appendix \ref{Appendix_rC} we show that
this specific value is given by
\begin{align}
\lamdimless^{(max)} \equiv  \frac{2}{3} \left(1 - \ln \frac{3\sqrt{2}}{\mu \rMh}  \right),
    \label{eq_chioutmax}
\end{align} 
We then denote deviations from this extreme value as $\Delta$, so that a general value for $\lamdimless $ is determined from
\begin{align}
 \lamdimless = \lamdimlessmax  + \Delta = \frac{2}{3} \left(1 - \ln \frac{3\sqrt{2}}{\mu \rMh}  \right) + \Delta,
    \label{eq_chiout_Delta}
\end{align}
thus, the actual value for the boundary condition is fully determined from $\Delta$.

We then determine (appendix \ref{Appendix_rC}) a better  estimate for $\rC$ which is
\begin{align}
\rC \approx& \frac{1}{3} \left(\frac{18 \rMh}{\mu^2} \frac{1}{ 1  + 3 |\Delta|  } \right)^{1/3}
    \label{eq:r_c}.
\end{align}
where the factor of $1/3$ is inserted to create a more conservative estimate.
Thus, our first naive estimate above ($\rC \approx \left(\frac{\rMh}{\mu^2}\right)^{1/3}$), simply corresponds to the maximum $\rC$ case which is obtained when $\Delta = 0$,
while the 2nd extreme case above is equivalent to $\Delta \gg 1$.  

 To summarize, $\rC$  denotes the scale where deviations from pure MOND solutions can occur due to the onset of the $\mu$ regime. 
As we show below, for distances $r > \rC$ the solution to the weak-field equations becomes oscillatory but still decaying (similarly to the Helmholtz equation).
Importantly, one may use data to constrain $\rC$, leading to global constraints on $\mu$ and specific constraints for $\Delta$ which may be different for individual astrophysical sources.
We discuss the dependence of the solution on $\Delta$ in section \ref{section:boundary_conditions}. 

\section{Numerical solutions: setting up the problem}
\label{Spher_sym}

In order to investigate how {\AEST}~theory solutions differ from those of the classical MOND theory, we define a baryonic system representative of an idealized spherical galaxy and calculate the solutions of the field equations, as described in the previous section.  In the case of MOND, solutions for the force for spherical 
systems can be obtained analytically starting from a Newtonian solution. However, due to the mass term $\mu$,
  analytical solutions are no longer possible in \AEST~theory
 and we need to rely on a numerical approach.  We describe in this section the setup that we use to obtain the spherically symmetric solutions.

\subsection{Density profiles}

We use two different density profiles for the analysis: a top-hat and a Hernquist profile, the latter being a good description 
of the baryonic component of spherical galaxies, see \cite{Hernquist_1990}. 

The top-hat profile is defined as a sphere of radius $a_h$ of constant density embedded in a uniform background:
\begin{equation}
\label{definition_top_hat}
\rho_b(r) = \begin{cases}
  \rho_c & \text{$r < a_h$}, \\
  0 & \text{$r > a_h$},
  \end{cases}
\end{equation}
where $\rho_c$ is the baryonic density at the centre of the profile.  For our purposes, we fix the parameters of the profile to $\rho_c = 3.45\times 10^9 \Msolar \kpc^{-3}$ 
which is $\sim 10^7$ larger than the critical density of the Universe, and $\ah=2~\kpc$. This is representative of a galactic system. 
We discuss how changing the density affects the solutions in Section~\ref{section:central_density}.

The Hernquist density profile is given by \citep{Hernquist_1990},
\begin{equation}
    \rho_b(r)=\frac{\rho_c}{\frac{r}{a_h}\left(1+\frac{r}{a_h}\right)^3}
    \label{eq:H_Density}.
\end{equation}
Note that for simplicity, we use the same symbol $a_h$ to represent the radius of the top-hat sphere as well as the scale radius for the Hernquist profile. The Newtonian potential associated to this profile is given by, \cite{Binney_Tremaine_2008}
\begin{equation}
    \Phi_N(r)=-\frac{2 \pi \GN \rho_c a_h^2}{1 + \frac{r}{a_h}}.
    \label{H_PhiN}
\end{equation}

\begin{figure}
    \includegraphics[width=0.9\columnwidth]{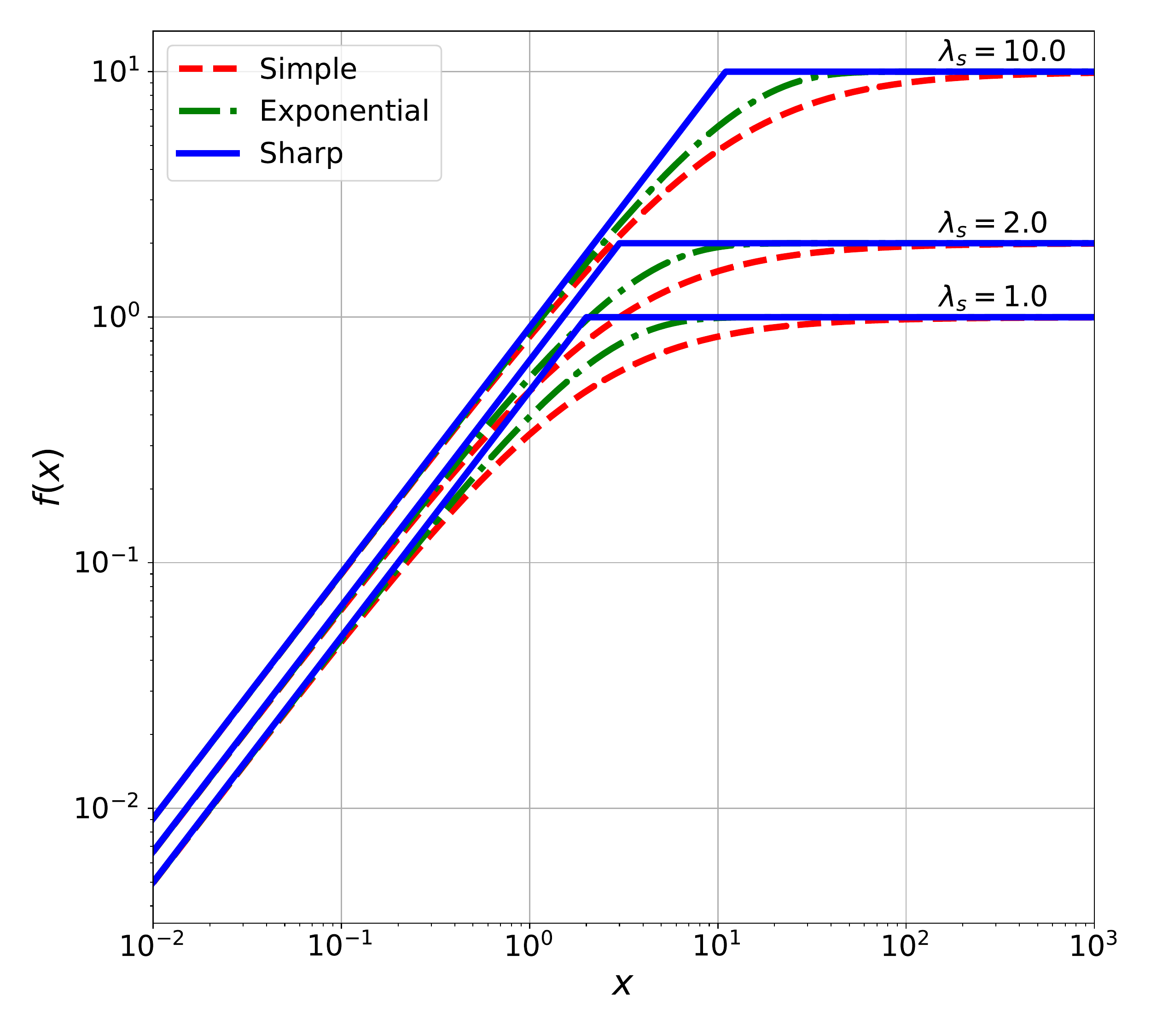}
    \caption{Interpolation functions explored in this work. Different colours correspond to the different functions defined by \eqref{mu_Simple}, \eqref{mu_Exp} and \eqref{mu_Hyp}.
    For each function, we show the result for three different values of $\lambda_s$: 1, 2 and 10.} 
    \label{fig:interpolation_functions}
\end{figure}

\subsection{{\AEST}~field equations for spherical isolated sources}

In this section, we focus on solving \eqref{Phi-Orig} and \eqref{Chi-AQUAL} in order to predict the radial dependence of $\Phih$ and $\chi$ which are critical for understanding
 the behaviour of gravity for spherical isolated sources in {\AEST}~theory.
The  potential $\Phi$ used for particle accelerations is calculated from \eqref{Phih_Phi_chi}.  The spherical version of \eqref{Phi-Orig} and \eqref{Chi-AQUAL}
 from which we obtained the numerical solutions is
\begin{align}
    \frac{d^2\tilde{\Phi}}{dr^2} &= \frac{4 \pi \GN}{1+\beta_0} \rho_b - \mu^2(\tilde{\Phi} + \chi) - \frac{2}{r}\frac{d \tilde{\Phi}}{dr},
\label{d2_Phi_spher_sym}
\\
\left[x\frac{df}{dx}+f \right]    \frac{d^2\chi}{dr^2} &= \frac{4 \pi \GN}{1+\beta_0} \rho_b - \mu^2(\tilde{\Phi} + \chi) - \frac{2}{r}f\frac{d\chi}{dr}, 
\label{d2_chi_spher_sym}
\end{align}
where $x=|\nabla\chi|/ a_0$. For setting boundary conditions we  employ the analytic solutions of these equations at the (non-zero) radius where we start the numerical integration. 
Throughout this work, we use $a_0 = 1.2 \times 10^{-10}$ m s$^{-2}$~\citep{Begeman_Broeils_Sanders_1991}.

We will use different expressions for the function $f$. Two are given in the literature \citep{2005MNRAS.363..603F, 2006ApJ...638L...9Z}:
\begin{align}
    \label{mu_Simple}
    & \text{Simple} &f(x) &= \frac{x}{1 + \beta_0 + \beta_0x} \\
    \label{mu_Exp}
    & \text{Exponential} &f(x) &= \lambdas \left(1 - e^{-\frac{x}{1 + \lambdas}}\right),  
\end{align}
which agree with the limits defined in \eqref{def_f_general}.  However, we will also explore the consequences of a different function that we define as
\begin{align}
    \label{mu_Hyp}
    &\text{Sharp} &f(x) &= \frac{x + (\lambda_s + 1) - |x - (\lambda_s + 1)|}{2 (1 + \beta_0)},
\end{align}
which exhibits a sharp Newtonian to MOND transition and returns the correct limits defined in \eqref{def_f_general}. This function has a turning point at $x\rightarrow \lambdas + 1$ and is not designed to be fully consistent with observations,  but rather as a test function to show the Newtonian to MOND transition radii and vice-versa.  Fig.\ref{fig:interpolation_functions} shows these three functions for three different values 
of $\lambda_s$.  See \cite{2008ApJ...683..137M} for additional discussion on interpolation functions.

\subsection{Analytic solutions for the top-hat profile}
\label{section:analytic_equations}

Analytic solutions are required  for setting up inner boundary conditions for the numerical solvers as well as for testing that our numerical implementation of the complete solutions is correct, at least for the cases where $\mu=0$ or $r \ll r_C$, for which these analytic solutions exist.  While  our numerical solutions are
obtained  using  \eqref{d2_Phi_spher_sym} and \eqref{d2_chi_spher_sym}, 
the analytical calculations are easier to obtain using \eqref{d2_Phi_spher_sym}  together with  the spherical version of \eqref{Chi-Orig}.

We now derive the analytic solutions for $\tilde{\Phi}$ and $\chi$ in the case of a top-hat source profile with the simple interpolation function.
In appendix \ref{Appendix_Sharp} we present solutions for the top-hat profile with the sharp interpolation function, while in appendix \ref{section:analytic_equations_Hernquist} 
we display the solution for the Hernquist profile
with the simple interpolation function.

The solution for $\tilde{\Phi}$ can be obtained by re-scaling the gravitational constant in the solution of the standard Poisson's equation \citep[e.g.][]{Binney_Tremaine_2008} and takes the following form:
\begin{align}
\Phit &=
    \begin{cases}
        - \frac{2 \pi \GN}{1+\beta_0} \rho_c (\ah^2 - \frac{r^2}{3}) & {\rm for }  \quad r < \ah 
        \\
        -\frac{4 \pi \GN}{3(1+\beta_0)} \rho_c \frac{\ah^3}{r} & {\rm for } \quad r > \ah
    \end{cases}, 
    \label{eq:tilde_Phi}
\end{align}
\begin{align}
    |\grad\tilde\Phi| &=
    \begin{cases}
        \frac{4 \pi \GN }{3(1+\beta_0)} \rho_c r & \text{ for } \quad r < \ah
        \\
        \frac{4 \pi \GN }{3(1+\beta_0)} \rho_c \frac{\ah^3}{r^2} & \text{ for } \quad r > \ah. 
    \end{cases}
    \label{eq:nabla_tilde_Phi}
\end{align}

We then integrate  \eqref{Chi-Orig} once to obtain
\begin{equation} 
    \grad\tilde\Phi = f (|\grad\chi|/a_0) \grad\chi  + \grad \times \mathbf{k}.
    \label{first_int}
\end{equation}

The field $\grad \times \mathbf{k}$ is divergenceless and was discussed in detail in \cite{1984ApJ...286....7B} where it has been shown to be exactly zero for particular symmetries (including spherical), and to behave at least as $r^{-3}$ for non-symmetric situations (the effects of this so called curl term on non-linear structure formation with pure MOND were studied in detail by \cite{Llinares:2008ce} and \cite{llinares_thesis}).  Since we are assuming spherical symmetry, we can ignore $\grad \times \mathbf{k}$  and invert equation \eqref{first_int} to find $\grad\chi$.  Applying this procedure with the simple interpolation function \eqref{mu_Simple} gives  
\begin{align}
    \grad\chi =&   \frac{\beta_0}{2}\left[1 +  \sqrt{1  + \frac{ 4 a_0 \lambdas (1 + \lambdas)}  {|\grad\tilde\Phi| } }  \right] \grad\tilde\Phi, 
    \label{eq:nabla_chi_simple}
\end{align}
which can be integrated to obtain a solution for $\chi$. 
Letting $M \equiv \frac{4}{3} \pi \ah^3 \rho_c$ be the total mass of the system, and defining the scale $\rint \equiv  4 \ah^3  / \rchi^2 $, the solution is
\begin{strip}
    \begin{align}
        \chi =&  
        \begin{cases}
\sqrt{  \GN M  a_0}  \left\{
		 \chiin
 + \frac{r^2}{\rchi \rint} 
+ \frac{1}{2}  \frac{ \sqrt{\rint r}}{\rchi}
 \sqrt{1 + \frac{r}{\rint}} \left( 1 + 2 \frac{r}{\rint}\right) 
		- \frac{1}{4} \frac{\rint}{\rchi} 
 \ln \left[1 + 2\sqrt{\frac{ r}{\rint}} \left( \sqrt{\frac{ r}{\rint}}   +    \sqrt{1 + \frac{r}{\rint}}  \right) \right] 
\right\}
& {\rm for } \; r \le \ah 
\\
		-  \frac{\GN M}{2 (1+\lambdas) r} 
    +  \sqrt{\GN M a_0} \left\{ \chiout - \sqrt{ 1 + \frac{\rchi^2}{4 r^2}   } 
+   \ln  \frac{  r }  { \rMh }   
+ \frac{1}{ 2 }  \ln    \left[ 1 + \frac{\rchi^2 }{ 8 r^2 }    +  \sqrt{ 1 + \frac{\rchi^2 }{ 4   r^2 }  }  \right]
 \right\}
& {\rm for } \; r \ge \ah  
\label{Simple_solution}
        \end{cases}
    \end{align}
\end{strip}

\noindent where $\chiin =  \chi(0)/ \sqrt{  \GN M  a_0} $ and  $\chiout$ are two integration constants corresponding to the inner and outer regions respectively, both normalized to $ \sqrt{  \GN M  a_0}$ for later convenience. 
These two integration constants are not independent but are related by matching the solution at the boundary $r=\ah$ in \eqref{Simple_solution}.
Lastly, we start our numerical integration, at radius $r_0$ (in the interior) and  set the boundary
condition as $\chi_0 = \chi(r_0) / \sqrt{  \GN M  a_0}$, which by appropriate use of \eqref{Simple_solution} may then be related to $\chiin$ and $\chiout$.
We study the dependence of the solution on the boundary condition in Section~\ref{section:boundary_conditions}.

\begin{figure*}
    \includegraphics[width=\textwidth]{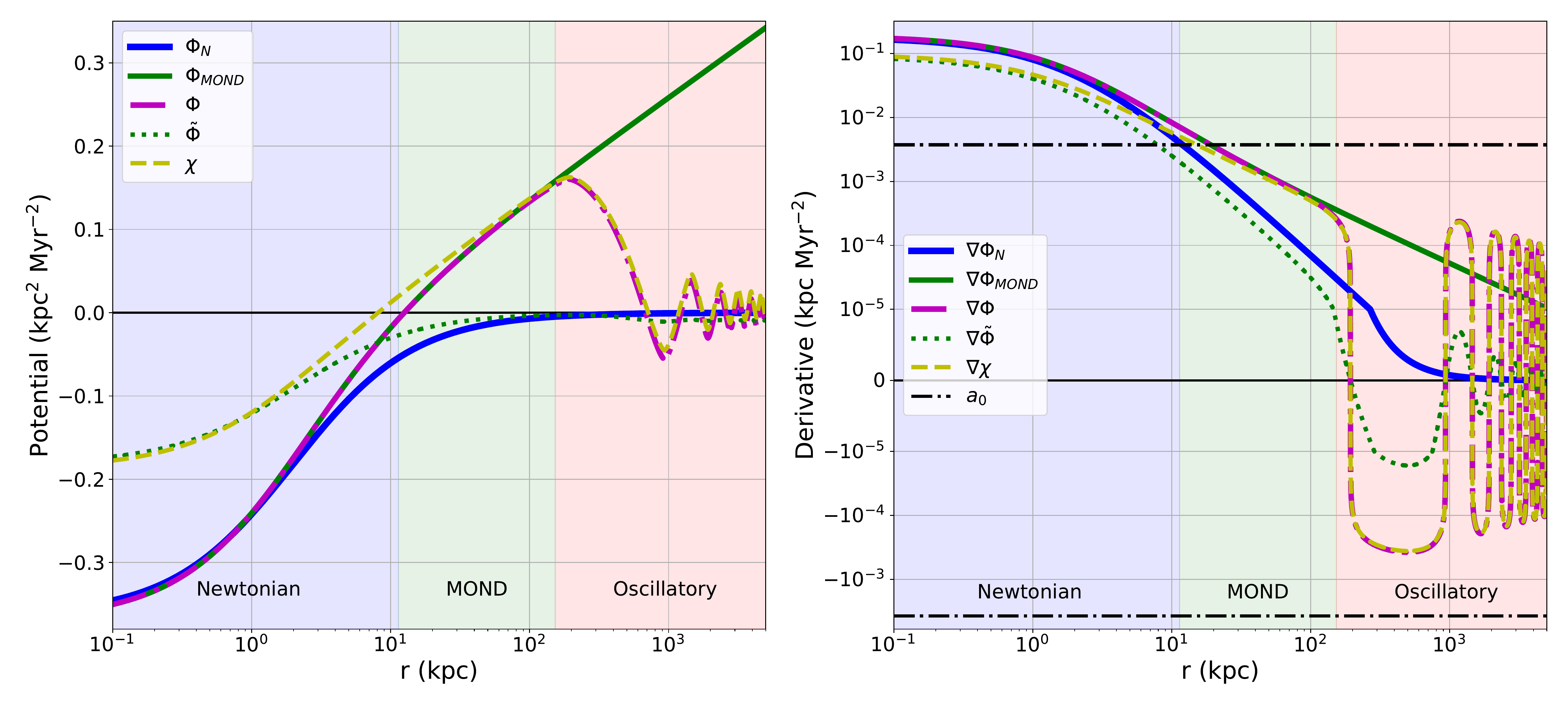}
    \caption{Numerical solution for the potentials (left) and their derivatives (right) for the Hernquist density profile and the fiducial model parameters with $(\lambda_s, \mu) = (1, 1~\Mpc^{-1})$.  The blue, green and red regions delineate the Newtonian, MOND and Oscillatory regions respectively. The yellow and green dashed lines are for the potentials $\tilde{\Phi}$ and $\chi$ respectively, and the purple dash-dot line is the potential $\Phi$ which is responsible for defining particle accelerations through its derivative.  We have included the Newtonian (blue) and classical MOND (green) solutions for comparison. The break in the blue curve at $\nabla\Phi=10^{-5}$ is not physical, but related to the symlog scaling that we use for the vertical axis of the right panel.}
    \label{fig:NS_l1m1_HQ_Simple}.
\end{figure*}


\begin{figure*}
    \includegraphics[width=0.75\textwidth]{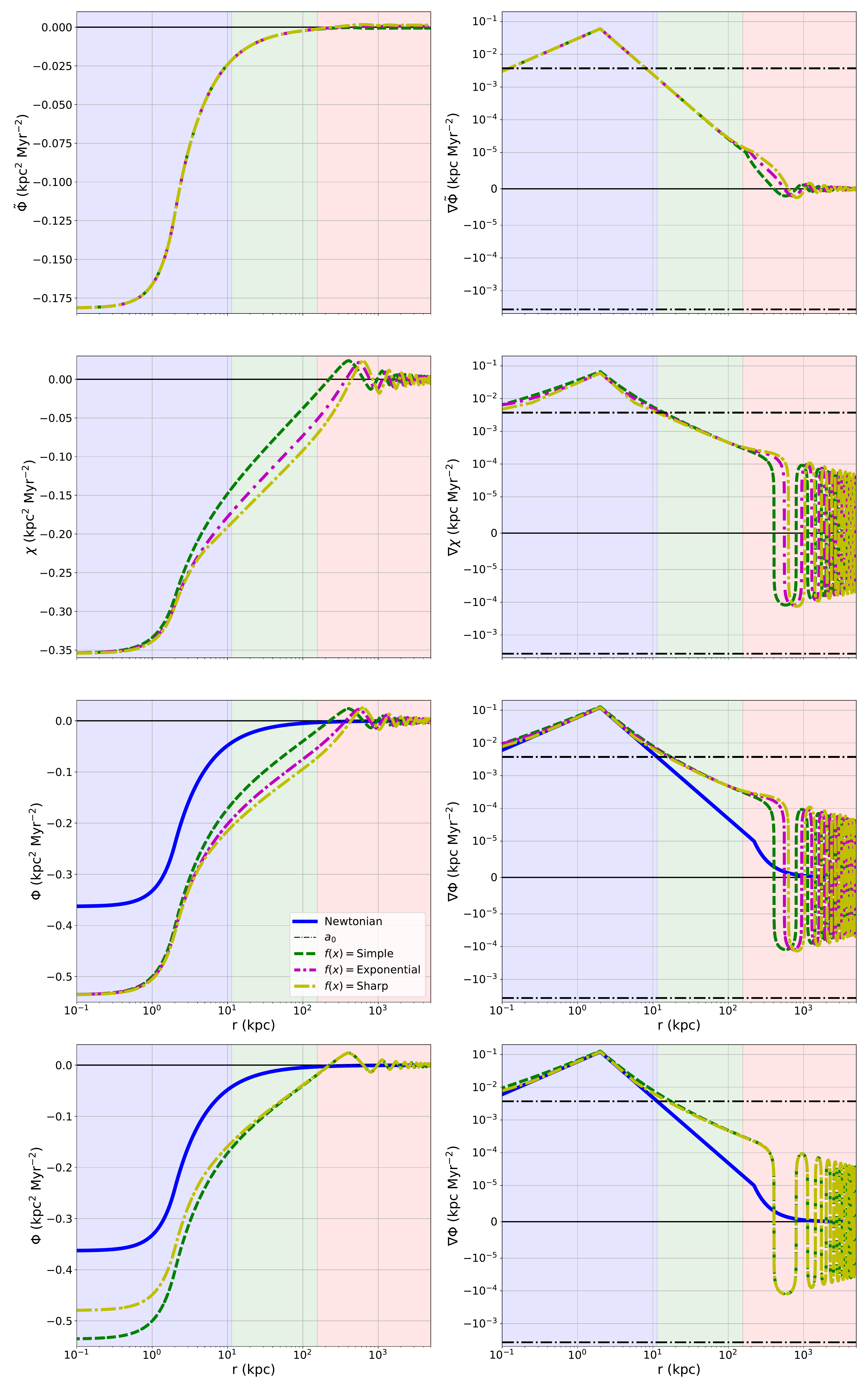}
    \caption{Sensitivity of the solutions to changes on the interpolation function $f$ for the top-hat profile with $(\lambda_s, \mu)=(1, 1\Mpc^{-1})$.  On the left we show the potentials and on the right their derivatives.   From top to bottom, the panels contain  $\tilde{\Phi}$, $\chi$ and $\Phi$.  The horizontal dash-dot lines in the right panels denote Milgrom's constant $a_0$. For each panel, the three lines each show an interpolation function and we also show the Newtonian solution in blue for comparison. The break in the blue curve at $\nabla\Phi=10^{-5}$ is related to the symlog scaling that we use for the vertical axis of the right panels. 
    }
    \label{fig:NS_l1m1_all_free_funcs}
\end{figure*}

\begin{figure*}
    \includegraphics[width=\textwidth]{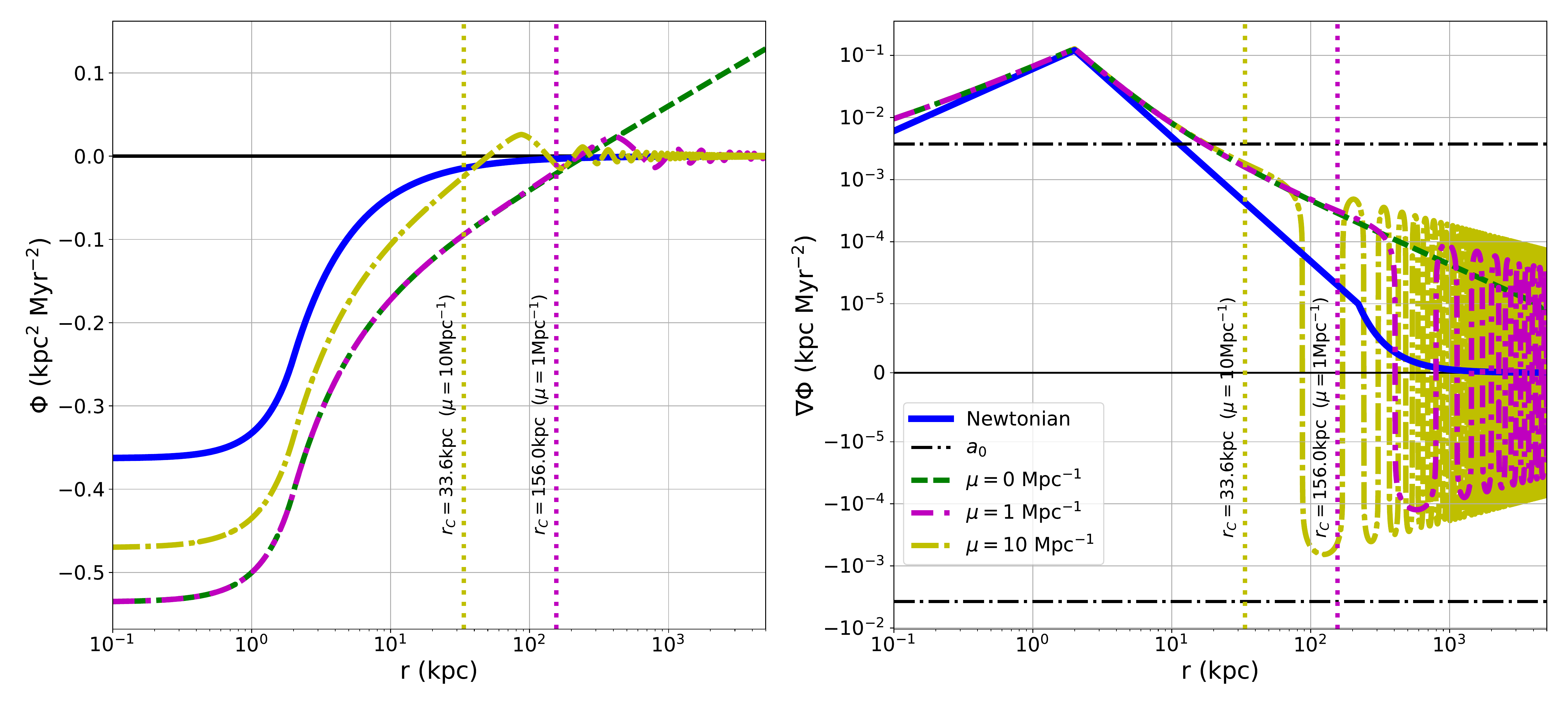}
    \caption{Sensitivity of the solutions to changes on the mass parameter $\mu$ on the potential $\Phi$ (Left) and its derivative (Right) for the top-hat profile and $\lambda_s=1$. The horizontal dash-dot lines in the right panels denote Milgrom's constant $a_0$.  The blue solid line is the Newtonian solution, shown for comparison.  The other three lines correspond to three different values of $\mu =\{0, 1, 10 \} \Mpc^{-1}$. The break in the curves at $\nabla\Phi=10^{-5}$ is related to the symlog scaling that we use for the vertical axis of the right panels.
    }
    \label{fig:NS_l1m-_TH_Simple}
\end{figure*}

\section{Results}
\label{results}

We first analyse the dependence of the numerical solutions on the model and density profile parameters. We then present a list of physical effects which make \AEST~theory different from standard MOND.

\subsection{General properties of the solutions}

Fig.\ref{fig:NS_l1m1_HQ_Simple} shows the numerical solution for the potentials (left) and their derivatives (right) for a fiducial set of parameters $(\lambda_s, \mu) = (1, 1 ~ \Mpc^{-1})$, 
the ``simple''  interpolation function $f(x)$ from~\eqref{mu_Simple} and the Hernquist density profile defined in~\eqref{eq:H_Density}.  
The blue solid curve in Fig.\ref{fig:NS_l1m1_HQ_Simple} corresponds to the Newtonian solution for $\Phi$ which we denote as $\Phi_N$ and the green solid curve is solution 
for $\Phi$ in {\AEST}~theory with $\mu=0$ 
which is equivalent to MOND, thus denoted as $\Phi_{\rm MOND}$.  Both $\grad \Phi_N$ and $\grad \Phi_{\rm MOND}$ agree in the central region of the galaxy 
and thus give rise to the same force profile there.  Farther away from the center, the gradients fall below $a_0$ (see horizontal lines in the right panel) 
 leading to the characteristic logarithmic MOND potential for $\Phi_{\rm MOND}$ outside the source and a force that follows a $1/r$ relation. 

The three additional curves in Fig.\ref{fig:NS_l1m1_HQ_Simple} in both panels are solutions of the field equations \eqref{d2_Phi_spher_sym} and \eqref{d2_chi_spher_sym} and the 
total potential provided by the relation \eqref{Phih_Phi_chi}.  The green dotted curve is the solution for $\Phit$ (and $\grad \Phit$); the yellow dashed curves are the solution for $\chi$ (and $\grad \chi$); 
and finally, the pink dash-dot curve is the solutions for the potential $\Phi$ (and $\grad \Phi$). 
 As we move farther away from the center, between the MOND radius $\rM$ and $\rC$, the theory tends to the classical MOND behaviour (i.e. a force law which not only has the same dependence with $r$, but also the same normalization). Farther away when $r>r_C$, the solutions enter the oscillatory regime, where the potential develops additional potential wells and the force can become repulsive.

In the next two sections, we  describe how variations of the \AEST~theory parameters affect these reference solutions and what additional physical effects are associated with them.
\begin{figure}
    \includegraphics[width=\columnwidth]{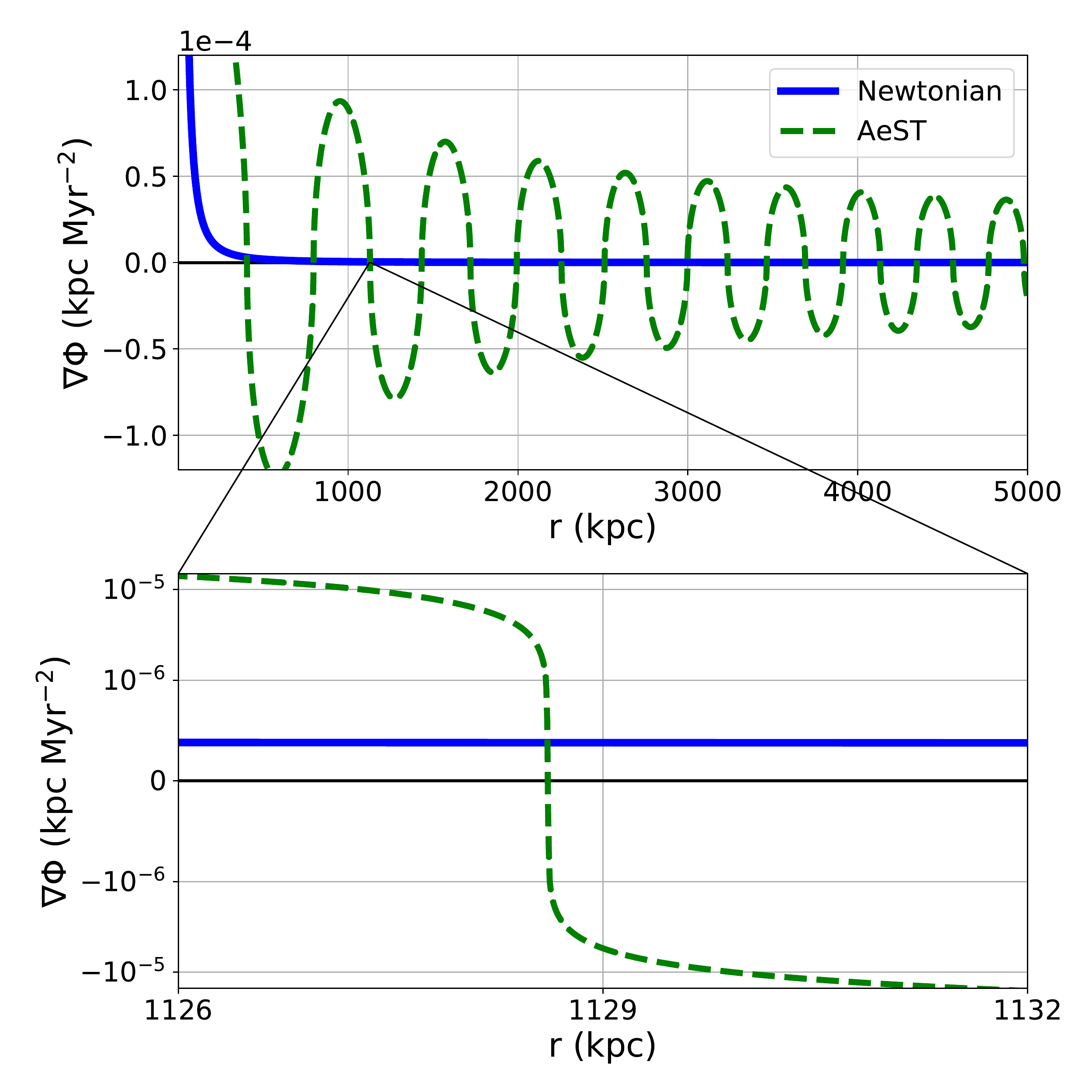}
    \caption{Spatial oscillations in the force profile for a model with $(\lambda_s, \mu) = (1, 1 \Mpc^{-1})$. Top plot is shown using a linear scaling for both axes. The bottom plot shows the detail of a zero crossing which exhibits a very steep slope. 
    }
    \label{fig:simple_linear_osc_detail}
\end{figure}

\begin{figure*}
    \includegraphics[width=\textwidth]{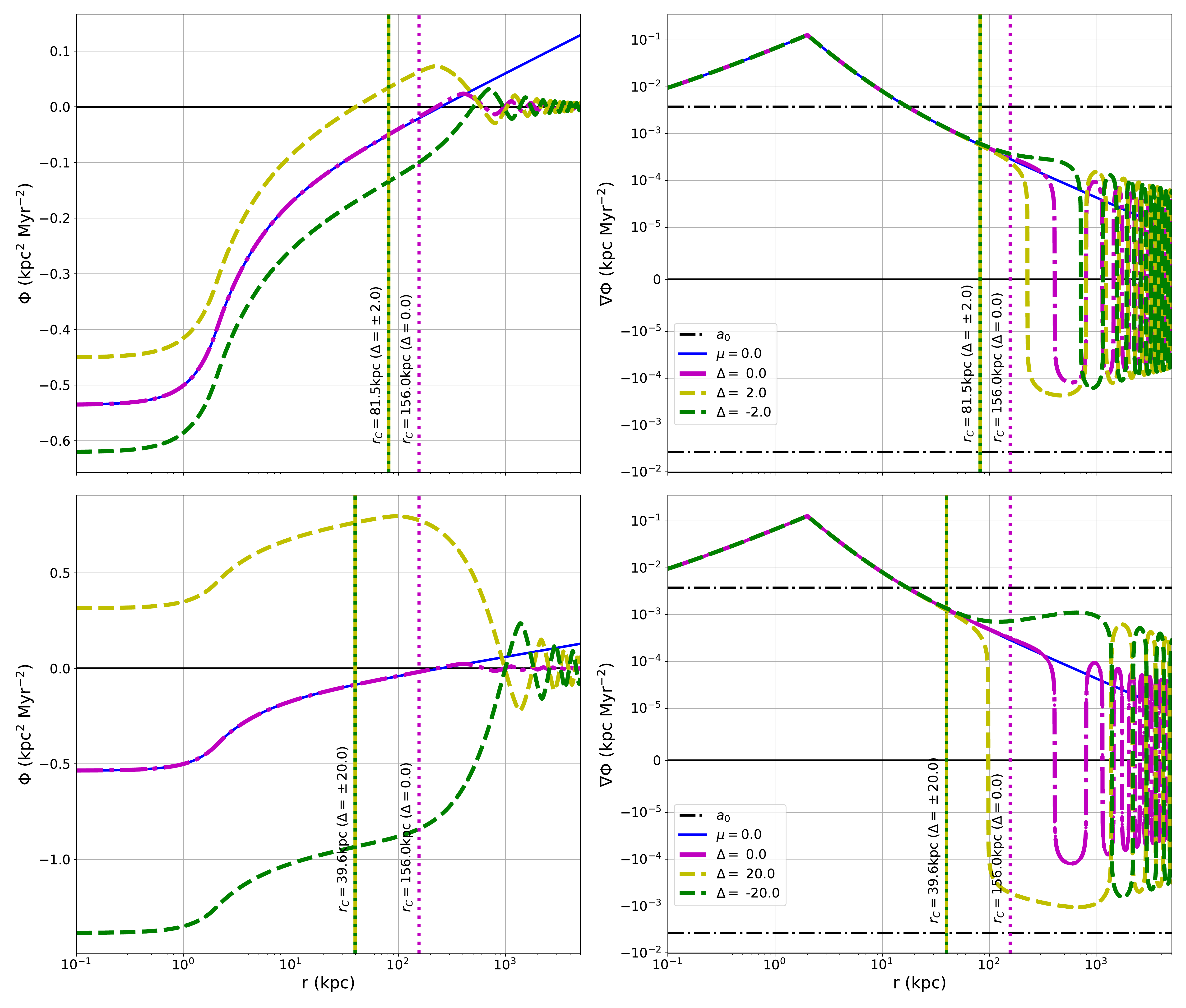}
    \caption{Sensitivity of the solutions to changes on the inner boundary condition employed to obtain numerical solutions.  We show solutions for metric perturbation $\Phi$ for the top-hat profile and $(\lambda_s, \mu)=(1, 1 \Mpc^{-1})$.  Left and right columns are the field and its radial derivative (i.e. the force that defines particle trajectories).  The horizontal dash-dot lines in the right panels are the MOND constant $a_0$.  The two different rows correspond to mild (upper row) and large (bottom row) offsets on the fiducial boundary condition which we define as the value of the Newtonian solution in the center.  The blue line is the Newtonian solution, shown for comparison. The break in the blue curve at $\nabla\Phi=10^{-5}$ is related to the symlog scaling that we use for the vertical axis of the right panels.} 
    \label{fig:NS_l1m1_TH_offset_Simple}
\end{figure*}

\subsection{Dependence on the free parameters}

We now study the dependence of the solutions on the free parameters of the model and on the central density (and mass) of the source, 
 assuming a top-hat density profile~\eqref{definition_top_hat}.

\subsubsection{Interpolation function}

Fig.\ref{fig:NS_l1m1_all_free_funcs} shows solutions for the gravitational potential (left column) and force (right column) that were obtained with
 the three interpolation functions $f(x)$ defined by \eqref{mu_Simple}, \eqref{mu_Exp} and \eqref{mu_Hyp}.  The first three rows correspond to a 
different potential; from top to bottom, these are $\tilde{\Phi}$, $\chi$ and $\Phi$.
We fix the boundary condition $\chi(r_0)$ at $r_0$ to be the same for all three interpolation functions, hence, the inner Newtonian regime is identical 
in all cases displayed in the first three rows of Fig.\ref{fig:NS_l1m1_all_free_funcs}. For the same reason, 
the MOND regime for the forces (right panels of Fig.\ref{fig:NS_l1m1_all_free_funcs}) is also identical in all solutions, however,
 the potentials (left panels of Fig.\ref{fig:NS_l1m1_all_free_funcs}) differ.
 The reason for the potentials reaching different values as the system evolves towards the MOND regime, is the sharpness or smoothness 
of the interpolation function $f(x)$. Smooth transitions from the Newtonian to the MOND regime lead to higher values of the potential before 
it settles into the MOND track (see also Fig.\ref{fig:interpolation_functions}), after which, its evolution becomes the same for all interpolation functions. 
When $r\gtrsim \rC$, the solutions enter the oscillatory regime, and the value of $\Phi=\Phit + \chi$ then plays a role due to the 
mass term $\mu$ (see \eqref{d2_Phi_spher_sym} and \eqref{d2_chi_spher_sym}). At that stage, since the value of the potential between interpolation functions is 
already different, the result is a change in the oscillation pattern, although the overall oscillation envelope stays the roughly the same; see Fig.\ref{fig:NS_l1m1_all_free_funcs}. 

The last row of Fig.\ref{fig:NS_l1m1_all_free_funcs} shows the result when the boundary conditions are calculated individually for each interpolation function
according to the prescription described in section~\ref{section:boundary_conditions}, as well as in Appendix~\ref{Appendix_rC} and \ref{Appendix_Sharp}.
With the boundary conditions appropriately calculated, the $\mu$-dominated region shows identical behaviour across all the interpolation functions shown. This is expected,
 as all interpolation functions have the same limiting form at small gradients. However, given that they now have different boundary conditions at $r_0$,  the inner Newtonian (and partially the transition
 to the MOND regime) is slightly different across each function.
We do not show the exponential interpolation function since we do not have an analytical solution for it in order to specify the $\chi$ boundary condition according to our prescription
of section \ref{Mond_to_mu} and Appendix~\ref{Appendix_rC}.

For the remainder of this work will mainly focus on the "Simple" interpolation function \eqref{mu_Simple} so as to simplify the plots and standardise the results.

\subsubsection{Mass parameter \texorpdfstring{$\mu$}{\emph{u}}}
\label{section:scalar_mass}

Setting the mass parameter $\mu$ to zero gives a standard MOND-like behaviour in which the force follows the Newtonian solution in the inner regions at $r<\rM$ and transitions into a MOND regime when $r>\rM$.  Letting $\mu$ take positive values initiates spatial oscillations in the solutions for $r \gtrsim r_C$.  Choosing $\lambdas=1$ and top-hat profile (leading to $\rM= 12.5 \, \kpc$), Fig.\ref{fig:NS_l1m-_TH_Simple} shows solutions with three different values of $\mu = \{ 0, 1, 10 \} \, \Mpc^{-1}$, corresponding to $\rC$: $\{\infty, 156, 33.6\} \, \kpc$; see \eqref{eq:r_c}. 
All solutions for the force give the same result up to $\sim \rC$. After this point, the solutions deviate from each other with the oscillations in the $\mu$-dominated region starting at smaller $r$ for larger $\mu$.

Observations of galaxies do not show evidence for the presence of repulsive gravitational forces affecting their internal dynamics, but only evidence of the MOND regime. 
Requiring $\rC$ to be larger than the virial radius of the Milky Way ($\sim 200 \, \kpc$) gives an estimate that $\mu^{-1} \gtrsim 1 \, \Mpc$. 
Parameter estimation using several different astrophysical objects and proper modelling of the observable quantities may
 provide more accurate bounds on this parameter but this goes beyond the scope of our current work. See however  \citep{Mistele:2023paq} where the first attempt
in doing so is considered.

Fig.\ref{fig:simple_linear_osc_detail} shows Newtonian and \AEST~forces for the region $r \gtrsim r_C$  assuming $\mu=1 ~\Mpc^{-1}$ and $\lambdas=1$.
Thanks to the non-linearity of the \AEST~theory equations, the oscillations are far from sinusoidal, acquiring rather a very steep slope when crossing zero. 
We generally find that the mean spatial frequency of the oscillations decreases with $\mu$, however, given the non-linear nature of the equations, we do not have an analytical estimate for the exact relation. 
Furthermore, the wavelength is not constant through the domain, but decreases slightly with $r$.  
Further detail of the zero crossing is shown in the bottom panel of Fig.\ref{fig:simple_linear_osc_detail} indicating that the 2nd derivative might become singular at the crossing. 
This, however, is an artifact, and an alternative way of solving the system of equations \eqref{Phi-Orig} and \eqref{Chi-AQUAL} using a Hamiltonian approach~\citep{DurakovicSkordis2023} shows that nothing bad happens there. 

\begin{figure*}
    \includegraphics[width=\textwidth]{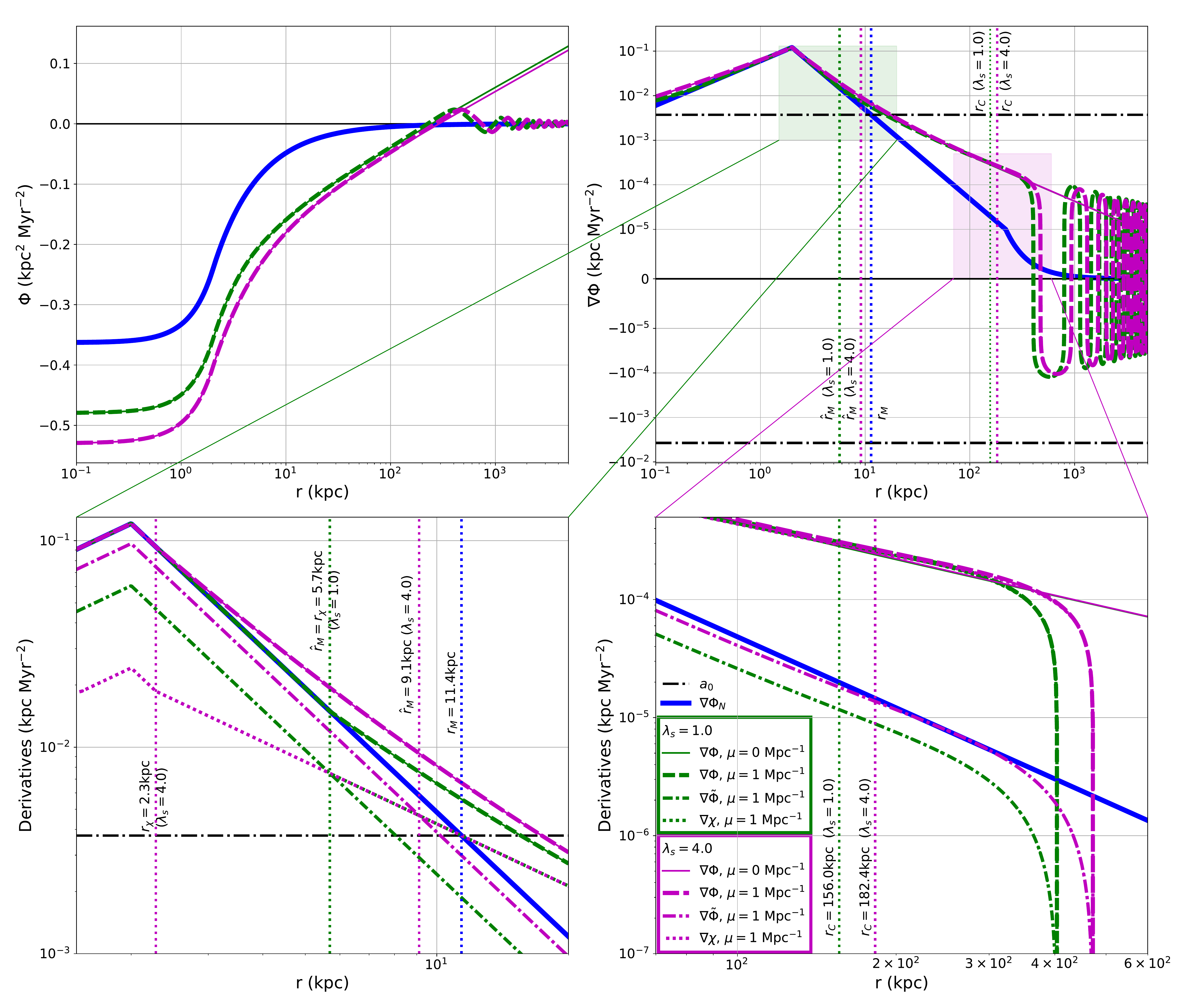}
    \caption{Sensitivity of the solutions to changes on the model parameter $\lambda_s$.  We show solutions for metric perturbation $\Phi$ for the top-hat profile and $\mu = 1 \Mpc^{-1}$.  In this particular figure we used the Sharp interpolation function to highlight the transition between the Newtonian and MOND limits. Top left and right plots are the field and its radial derivative (i.e. the force that defines particle trajectories).  The horizontal dash-dot lines are the MOND constant $a_0$. The blue line is the Newtonian solution, shown for comparison. The break in the blue curve at $\nabla\Phi=10^{-5}$ is related to the symlog scaling that we use for the vertical axis of the top right panel. See section \ref{section:lambda_s} for the explanation of the bottom panels.}
    \label{fig:NS_l-m1_TH_Hyp}
\end{figure*}

\begin{figure*}
    \includegraphics[width=\textwidth]{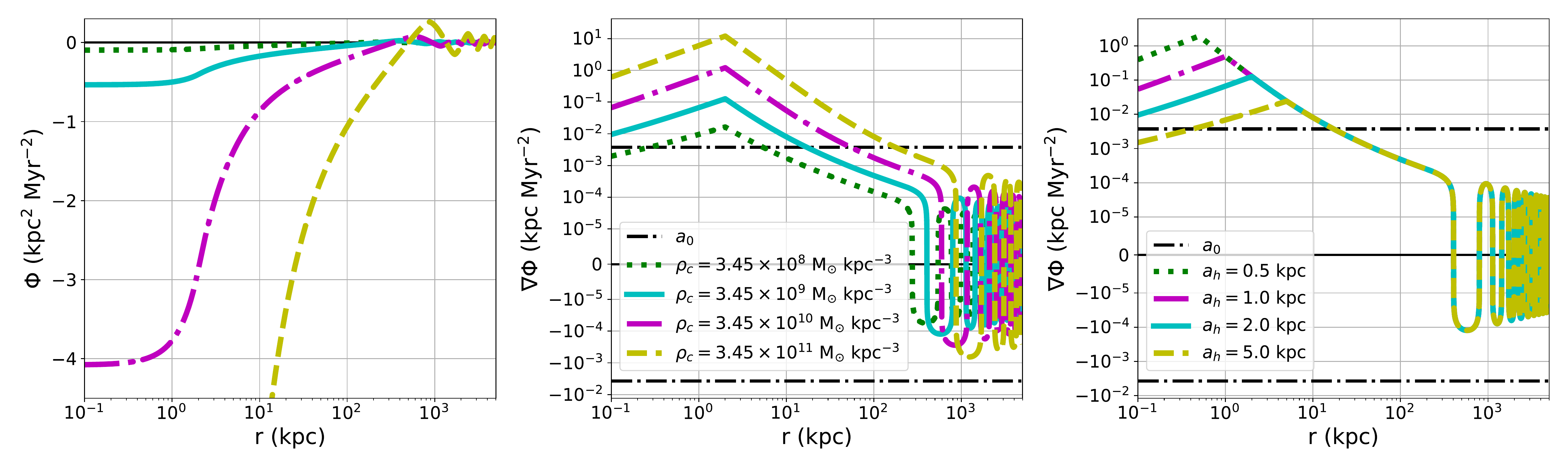}
    \caption{Sensitivity of the solutions to changes on the normalization of the density profile. The left panel shows the total potential $\Phi$ and the middle and right panels shows its 
radial derivative.  While in the left and middle panel the profile radius $\ah$ is kept constant (so that the total mass is different), in the right panel the mass is kept the same for all curves by 
adjusting $\ah$.  The continuous cyan curve is the fiducial model used in previous figures with $\rho_c=3.45\times 10^9~M_{\odot}\kpc^{-3}$. 
 The other curves show perturbations around this value.  The inner boundary condition for the potential is the same as used in previous figures.} 
    \label{fig:NS_d-l1m1_TH_Simple} 
\end{figure*}

\subsubsection{Boundary conditions}
\label{section:boundary_conditions}

The Poisson and classical MOND equations depend on the potential only through its derivatives, hence, adding a constant to the solution is not physically relevant and does not affect the gravitational force.  
The situation is different in {\AEST}~theory  due to the presence of the mass term $\mu^2\Phi$ in \eqref{d2_Phi_spher_sym} and \eqref{d2_chi_spher_sym} 
which makes the value of the potentials physically relevant at $r>\rC$.
In other words, the boundary conditions are important in \AEST~theory. 
We  now investigate their impact on the solution determining the depth and radial profile of the gravitational potential $\Phi$ and its derivative.

As discussed in Sec.\ref{Mond_to_mu} the boundary condition $\Delta$ directly affects where the solution departs from pure MOND and enters the $\mu$-regime. This is 
estimated according to \eqref{eq:r_c}.  The effect of varying $\Delta$ is illustrated in Fig.\ref{fig:NS_l1m1_TH_offset_Simple} and as the numerics confirm,
  $\Delta = 0$ constitutes the fiducial zero point corresponding to maximum $\rC$.

To investigate this, we choose a value $\chi_0$ at $r=r_0$, such that, $\chiout=  \lamdimlessmax  + 1 -  \frac{1}{2} \ln 2$ through the use of \eqref{Simple_solution}, 
 where $\lamdimlessmax$ is defined in \eqref{eq_chioutmax}.  
This corresponds to $\Delta = 0$ in \eqref{eq_chiout_Delta} as shown by the magenta dash-dot curve in Fig.\ref{fig:NS_l1m1_TH_offset_Simple}. 
The expectation from  \eqref{eq:r_c} is that this also corresponds to the maximum $\rC$ when all the other
parameters, e.g. $\mu$ and $\lambdas$ are kept fixed. Indeed, this is what is observed in the upper row of Fig.\ref{fig:NS_l1m1_TH_offset_Simple}. 
Varying $\Delta$ to negative (positive) values, results in $\chiout$ being smaller (larger). However, in both cases $\rC$, decreases from the value reached through $\Delta=0$, 
as expected through \eqref{eq:r_c} where $\Delta$ enters with an absolute value. 

There are two further related effects that are worth describing. Firstly, when $\Delta$ is positive, the force $\grad \Phi$ 
 passes through zero  at smaller distances than the fiducial $\Delta=0$ case. Hence, one observes a smaller oscillation phase. 
Conversely, for negative $\Delta$ the opposite happens, that is, $\grad \Phi$  passes through zero at larger distances than the fiducial $\Delta=0$ case. Hence, one observes  a larger oscillation phase.
This is seen in both upper and lower panels of  Fig.\ref{fig:NS_l1m1_TH_offset_Simple}.
Secondly, when $\Delta$ deviates significantly from zero, the force reaches an extensive plateau which is either negative (for $\Delta>0$) or positive (for $\Delta<0$) before entering the 
oscillatory regime.
This effect is more pronounced in the lower panel of Fig.\ref{fig:NS_l1m1_TH_offset_Simple}.

\subsubsection{\texorpdfstring{Parameter $\lambda_s$}{Screening Parameter}}
\label{section:lambda_s}

The screening parameter $\lambdas$ (and associated parameter $\beta_0=1/\lambdas$) primarily relates to the relative contribution of the $\chi$ field to the total field $\Phi$ in the deep Newtonian limit, see Eq.~\eqref{Jcal_GR}.
This parameter  must be positive \citep{Skordis:2021mry}, with $\lambdas \rightarrow \infty$ corresponding to the $\chi$ field being totally screened in the Newtonian limit (i.e. $\chi \rightarrow 0$). 

The $\lambdas$  parameter enters the definition of the following  scales
\begin{itemize}
    \item $\rint=4\ah^3/\rchi^2=4(1+\lambdas)^2\ah^3/\rM^2$  for the interior solution, 
    \item $\rchi = \rM/(1+\lambdas)$ and $\rMh = \rM/(1+\beta_0)$ for the MOND regime, 
    \item $\rC\propto (1+\beta_0)^{-1/3}$ for the transition to $\mu$-dominant oscillatory regime. 
\end{itemize}
In addition, $\lambdas$ controls the effective gravitational strength which couples $\Phit$ to the density $\rho$, see \eqref{d2_Phi_spher_sym} 
and also \eqref{eq:tilde_Phi} and \eqref{eq:nabla_tilde_Phi}.

Fig.\ref{fig:NS_l-m1_TH_Hyp} shows the affect of varying $\lambdas$ on the solution of the potential $\Phi$ whilst keeping other parameters fixed. We have used the Sharp interpolation function \eqref{mu_Hyp} to show these effects clearly.
The zoom-in portion of the figure (bottom left) shows the transition from Newtonian to MOND behaviour in the exterior of the source. We see the two scales $\rMh$ and $\rchi$ at work, and their dependence on $\lambdas$. The scale $\rchi$ signifies when the $\chi$ component, and the total field $\Phi = \chi+\Phit$, changes from a Newtonian to MOND behaviour. The scale $\rMh$ signifies the point at which $|\grad\chi|$ becomes the dominant component of the total force ($\propto |\grad\Phi| = |\grad\chi + \grad\Phit|$). The transition to the full MOND force is essentially completed when the point $\rM$ is reached.
We note that both Newtonian and MOND magnitudes of the force are independent of $\lambdas$ -- only the transition is affected by this parameter.

Towards larger radii, when $r\sim \rC$, the solution approaches the $\mu$-dominated regime as discussed above. There is a mild dependence on $\lambdas$ as to where that happens, since $\rC\propto \lambdas^{1/3}/(1+\lambdas)^{1/3}$, and this is clearly seen in the zoom-in region (bottom right). The main effect of this shift, is the starting point of the oscillations, which sets the overall oscillation phase.

\subsubsection{Central Density and total mass}
\label{section:central_density}

In our discussion so far, we have  assumed a fiducial central density representative of the bulge of the Milky Way or an average spherical galaxy~\citep{Widrow_2008}.  
In Fig.\ref{fig:NS_d-l1m1_TH_Simple} we show the effect of changing this central density, on the total potential $\Phi$ (left panel), on the total force $\grad\Phi$ (middle panel)
 and on the total force when the mass is kept constant (by adjusting the source size $\ah$) (right panel) 

We observe that larger central densities (resulting in larger total mass), make the inner potentials deeper and increase the overall force throughout the Newtonian and MOND regimes,
 but also increase the magnitude of the oscillations in the $\mu$-dominated regime. The opposite happens with decreasing the central density, 
and this overall behaviour is what we expect based on the transition scale dependence on the total mass $M$.  Indeed, in the right panel, we 
see the effect of keeping $M$ constant, which is that the exterior solution remains the same, and the only differences occur in the interior.

\section{Conclusions}

In this paper, we predict the profile of the gravitational potential and associated force expected in spherical galaxies in the {\AEST}~theory.  The later is the first extension of GR
that successfully fits the CMB angular and matter power spectra without a dark matter component.
The non-relativistic limit of {\AEST}~theory differs from the classical MOND theory in the fact that the field equations (equivalent to the Poisson's equation in the standard gravity case) include a mass term. 
 This new ingredient leads to a different gravitational phenomenology which we investigate here for two different spherical density distributions: a top-hat and a Hernquist profile.  

We identified three characteristic regimes in the solutions, independent of the density profile: a Newtonian regime, 
a MOND regime and finally, an oscillatory regime, where the mass term dominates and the fields (as well as the forces and their associated dynamical mass) develop spatial oscillations.  
Focusing on the case where the MOND regime appears for intermediate scales and the oscillatory regime at larger scales, we find that the transition from the Newtonian to the MOND regime 
depends on the usual acceleration parameter $a_0$, and the screening parameter $\lambda_s$, while the transition from MOND to the oscillatory regime depends on the mass parameter $\mu$, 
the inner boundary condition and the total mass of the gravitating object through \eqref{eq:r_c}.  This means that the oscillations do not appear at a fixed radius, but have a more 
complex dependence which is different for every object and depends on their mass distribution.
On the other hand, we find no strong dependence of the solutions with the free function that regulates the speed of the transition between the Newtonian and MOND regimes. 

There should be remarkable consequences from the distinct behaviour of the gravitational potential and forces observed here. Accurate predictions would require N-body simulations but we do expect the matter density and distribution to be impacted and different from that in the $\Lambda$CDM model. Given the oscillations at large scales corresponding to the $\mu$ dominated regime, one could speculate that this will translate into the existence of ring-like structures far away from the galactic centre, which may eventually resemble observed structures such as galactic rings \citep{1996FCPh...17...95B}. Furthermore, the enhancement (suppression) of the {\AEST}~gravitational potential in the inner galactic regions for certain values of the inner boundary condition (see Fig.\ref{fig:NS_l1m1_TH_offset_Simple}) 
and interpolation function (see Fig. \ref{fig:NS_l1m1_all_free_funcs}) could be misinterpreted as a higher (lower) dark matter density in the $\Lambda$CDM framework. 
This could affect lensing analyses, galaxy cluster profiles, dark matter indirect (and potentially also direct) detection predictions, 
and presumably the whole galactic (and possibly early Universe) evolution.       

The main conclusion of our work is that {\AEST}~theory is a potential alternative to particle dark matter but N-body simulations are needed to make sure that the non-linear formation of 
large-scale-structures in this theory and the gravitational structure of  galaxies are consistent with observations.

\section*{Acknowledgements}

We thank Amel Durakovic for discussions and David F. Mota for valuable input and insights that have improved the quality of this work. We thank Claudio Llinares for his contribution at an earlier stage of this project. This work was supported by the European Structural and Investment Fund and the Czech Ministry of Education, Youth and Sports (Project CoGraDS - CZ.02.1.01/0.0/0.0/15\_003/0000437).





\bibliography{References}


\appendix

\section{Derivation of $\rC$}
\label{Appendix_rC}

In this appendix we give a more accurate estimate for $\rC$, which is the scale for which the solution to the system of equations \eqref{d2_Phi_spher_sym} and \eqref{d2_chi_spher_sym} deviates from the pure MOND solution, due to the presence of the $\mu^2$ term. Our strategy is to take the MOND solution as a background, perturb around it, and determine when the perturbation leads to sizable deviations. 

In the deep MOND regime, the details of the interpolation function are unimportant and we may set $f\rightarrow x/ (1+\beta_0)$. Moreover,since in the deep MOND regime $\Phit \sim 1/r^2$ is always smaller than $\chi$ it is sufficient to truncate \eqref{d2_Phi_spher_sym} and \eqref{d2_chi_spher_sym}  into a single equation for $\chi$ as
\begin{align}
 \frac{1}{r^2} \frac{d}{dr} \left[  r^2 \left|\frac{d\chi}{dr}\right|  \frac{d\chi}{dr} \right]
  + (1+\beta_0) a_0 \mu^2 \chi
=0.
\end{align}

We let $\chi = \chib +  \sigma $ where $\sigma$ is a perturbation to the background exact MOND solution which for large $r$ maybe approximated by 
\begin{align}
\chib \approx  \sqrt{\GN M a_0} \left( \lamdimless +  \ln \frac{r}{\rMh} \right),
\label{chi_approx}
\end{align}
where $\lamdimless$ depends on the details of the interpolation function $f$. For example, in the case of the simple interpolation function given by \eqref{mu_Simple} we have that $\lamdimless =  \chiout   -  1 +  \frac{1}{2} \ln 2$ while in the pure MOND function (i.e. $f= x$ exactly) we have that  $\lamdimless$ is the a free parameter by itself.

The sign of $\chi$ is dominated by $\chib$ (by virtue of $\sigma$ being a perturbation) so that we obtain the following linear equation for $\sigma$:
\begin{align}
 \frac{1}{r^2} \frac{d}{dr} \left[  r   \frac{d\sigma}{dr} \right] +   \frac{\mu^2}{2\rMh} \sigma = -\frac{\mu^2}{2\rMh}  \chib
\end{align}

The solution is a linear combination of Bessel functions $J_0$ and $Y_0$ plus the particular integral  which turns out to be exactly $\chib$ plus $\frac{\pi}{3} \sqrt{\GN M a_0} Y_0$. Hence,the full solution is
\begin{align}
\frac{\chi}{\sqrt{\GN M a_0} }  =&  C_1 J_0\left(  \sqrt{   \frac{2\mu^2}{9\rMh}  }   r^{3/2} \right) + \frac{\pi}{3} C_2 Y_0\left(\sqrt{   \frac{2\mu^2}{9\rMh}  }   r^{3/2} \right)
\nonumber
\\
 & +  \frac{\pi}{3}  Y_0\left(\sqrt{   \frac{2\mu^2}{9\rMh}  }   r^{3/2} \right),
\end{align}
for constants $C_1$ and $C_2$.  Expanding the above solution around $\mu= 0$  we find that $\chi\rightarrow const + const\ln(r)$ which has the same form as \eqref{chi_approx}. We match \eqref{chi_approx} exactly, by appropriately choosing $C_1$ and also setting $C_2=0$, so that the $\chi$ solution for $r \lesssim \rC$ is
\begin{align}
\frac{\chi}{ \sqrt{\GN M a_0}} \approx&   \left(  \lamdimless   -  \frac{2}{3}\gamma_{EM} - \frac{2}{3} \ln\frac{\mu \; \rMh}{3\sqrt{2}}  \right) J_0\left(  \sqrt{   \frac{2\mu^2}{9\rMh}  }   r^{3/2} \right)
\nonumber
\\
&
\ \ \ \ +  \frac{\pi}{3}  \; Y_0\left(\sqrt{   \frac{2\mu^2}{9\rMh}  }   r^{3/2} \right),
\end{align}
where $\gamma_{EM}$ is the Euler-Mascheroni  constant. 

We now expand as a series in $\mu$ to get
\begin{align}
\frac{\chi}{ \sqrt{\GN M a_0}} \approx&    \lamdimless + \ln \frac{r}{\rMh} + \left[ \frac{2}{3} - \lamdimless - \ln \frac{r}{\rMh}\right] \frac{\mu^2 r^3}{18 \rMh}.
\label{chi_full_approx}
\end{align}
Using this expansion to get our $\rC$ estimate can be problematic  because the first term becomes zero at a distance $\rMh e^{-\lamdimless}$. Thus we estimate $\rC$ not for $\chi$ but for it's derivative $|\grad\chi|$, which still provides a good estimate for our purposes. Then
\begin{align}
\frac{|\grad\chi|}{ \sqrt{\GN M a_0}} \approx&     \frac{1}{r} - \left[ 3\ln \frac{r}{\rMh} +   3\lamdimless - 1\right] \frac{\mu^2 r^2}{18 \rMh},
\end{align}
so that the 1st term is manifestely positive. Our estimate is then found through equating the  2nd to the 1st term, leading to
\begin{align}
\left( 3\ln \frac{\rC}{\rMh} +   3\lamdimless  -1  \right) \frac{\mu^2  \rC^3}{18\rMh} = 1.
\label{xi_def_equation}
\end{align}

Assuming for the moment that $3\ln \frac{\rC}{\rMh} +   3\lamdimless  -1>0$, we define
\begin{align}
\zeta \equiv 3\ln \frac{\rC}{\rMh} +   3\lamdimless  -1,
\\
\zeta_0 \equiv \frac{18 e^{3\lamdimless -1} }{\mu^2 \rMh^2},
\end{align}
so that \eqref{xi_def_equation} becomes
\begin{align}
W(\zeta) \equiv \zeta e^\zeta  = \zeta_0,
\end{align}
where $W(\zeta)$ is the Lambert function, hence $\zeta(\zeta_0)$ defines a one-parameter family of solutions. While $\zeta(\zeta_0)$ must in general be determined numerically, typically $\zeta_0$ is huge, and then we may approximate the solution as
\begin{align}
e^\zeta\approx& \frac{\zeta_0}{\ln \zeta_0  },
\\
\approx&  \frac{18 e^{3\lamdimless -1} }{\mu^2 \rC^2 \left[ \ln \frac{18}{\mu^2 \rMh^2} + 3 \lamdimless -1  \right]}.
\end{align}
Subbing $\zeta$ then gives us
\begin{align}
  \rC  =  \left[\frac{18\rMh}{\mu^2  \left( \ln \frac{18}{\mu^2 \rMh^2} + 3 \lamdimless -1  \right)}  \right]^{1/3}.
\label{rC_est_1}
\end{align}

Our above estimate, however, only makes sense if $ 3\ln \frac{\rC}{\rMh}  +   3\lamdimless  -1 $ is positive. On the contrary it would appear that it may even vanish for certain (negative) $\lamdimless$ which would invalidate the estimate.

This behaviour does not happen. Rather, there exists a special value for $\lamdimless \rightarrow \lamdimlessmax$ for which $\rC$ is a maximum. This value is at the point for which the 2nd term in \eqref{chi_full_approx} reaches zero, that is, when

\begin{align}
 \lamdimlessmax = \frac{2}{3} - \ln \frac{\rC^{(max)}}{\rMh},
\end{align}
which is indeed negative.  We first estimate this $\rC^{(max)}$ by subbing $\lamdimlessmax$ into \eqref{xi_def_equation} to get

\begin{align}
  \rC^{(max)} = \left(\frac{18\rMh}{\mu^2} \right)^{1/3},
\end{align}
so that

\begin{align}
 \lamdimlessmax = \frac{2}{3} \left(1 - \ln \frac{3\sqrt{2}}{\mu \rMh}  \right).
\end{align}

Generally, we may set 
\begin{align}
 \lamdimless = \lamdimlessmax  + \Delta,
\end{align}
where $\Delta$ parametrizes deviations away from $\lamdimlessmax$. We don't use the above relation directly into \eqref{rC_est_1}. Rather, we resort to the fact that $\rC$ is bounded from above exactly at $\Delta=0$, and so we take its absolute value when used in \eqref{rC_est_1}. This then leads to
\begin{align}
\rC \approx& \frac{1}{3} \left(\frac{18 \rMh}{\mu^2} \frac{1}{ 1  + 3 |\Delta|  } \right)^{1/3},
\end{align}
which is devoid of any pathologies resorting from negative values. Notice that we inserted an additional factor of $1/3$ by hand in order to get a more conservative estimate.

\section{Analytic solutions for the sharp interpolation function}
\label{Appendix_Sharp}

For the sharp interpolation function is particularly simple to obtain analytical solutions for the $\mu=0$ case.
When $x\ge 1+\lambdas$ then $f = \lambdas$ so that $\grad\chi = \beta_0 \grad\Phit$ which can be easily integrated both in the interior and exterior case. 
When $x\le1+\lambdas$ then $x \grad\chi = (1+\beta_0) \grad\Phit$, which can also be easily integrated. The final solution is then
\begin{align}
\chi &=  \begin{cases} 
 \sqrt{ \GN M a_0}\left( \chiin^{(M)} + \frac{2}{3}   \frac{r^{3/2}}{\ah^{3/2}} \right) r \le \frac{1}{4} \rint
\\
  \sqrt{\GN M a_0}\left( \chiin +  \frac{2r^2}{\rint \rchi} \right)   \quad   \frac{1}{4} \rint  \le r \le \ah
\\
  \sqrt{\GN M a_0}\left( \chiout -  \frac{\rchi}{r}   \right)   \quad  \ah \le r \le \rchi
\\
  \sqrt{\GN M a_0} \left( \lamdimless + \ln\frac{r}{\rMh} \right)  \quad r \ge \rchi
\end{cases},
\end{align}
where the relations between the different boundary constants are
\begin{align}
\chiout =&  \lamdimless + 1 + \ln\frac{1+\beta_0}{1+\lambdas} ,
\\
\chiin =& \chiout  - \frac{3\rchi}{ 2 \ah} ,
\\
\chiin^{(M)} =& \chiin  -  \frac{1}{6} \frac{\ah^3}{\rchi^3} .
\end{align}
Note that the sharp transition  between $x\ge 1+\lambdas$  and $x\le 1+\lambdas$  occurs exactly at $r=\rchi$.

\section{Analytic solutions for the Hernquist profile}
\label{section:analytic_equations_Hernquist}

We briefly discuss analytic solutions in the case of the Hernquist profile. This are indeed possible to obtain in the case of the simple interpolation function. Following the same procedure as in section \ref{section:analytic_equations} we find that the interior solution ($r\ge \ah$) for $\chi$ is

\begin{align}
\chi =& -\frac{2 \GN M}{1+\lambdas} \frac{1 + \sqrt{1+ \left(\frac{\ah + r}{\rchi}\right)^2 }}{\ah + r}
 + \sqrt{a_0 \GN M} 
\nonumber
\\
& 
 \times \ln\left\{ 1 + 2 \left(\frac{\ah + r}{\rchi}\right)^2  \left[1 + \sqrt{1 + \left(\frac{\rchi}{\ah + r}\right)^2 }\right] \right\},
\end{align}

\noindent while the exterior solution remains the same as for the top-hat profile.


\bsp	
\label{lastpage}
\end{document}